\documentclass[preprint]{aastex}

\shorttitle{SOLAR CYCLE CHANGES IN GONG DATA}
\shortauthors{Jain, Tripathy, and Bhatnagar}

\begin{document}
\title{SOLAR CYCLE INDUCED VARIATIONS IN GONG {\it p}-MODE FREQUENCIES AND SPLITTINGS}
\author{ Kiran Jain,\altaffilmark{1} S. C. Tripathy,\altaffilmark{2} 
and A. Bhatnagar\altaffilmark{3}}
\affil{Udaipur Solar Observatory, Physical Research Laboratory\\ 
P.O. Box No.198, Udaipur - 313 004, INDIA \\}
\altaffiltext{1}{kiran@uso.ernet.in}
\altaffiltext{2}{sushant@uso.ernet.in}
\altaffiltext{3}{arvind@uso.ernet.in}

\begin{abstract} 
 We have analysed the recently available GONG {\it p}-mode frequencies and splitting
 coefficients for a period of three and half years, including the rapidly rising phase of
 solar cycle 23. 
 The analysis of mean frequency shift with different 
 activity indices shows that the shift is equally correlated with both  magnetic and 
 radiative indices.  During the onset of the new cycle 23, we notice that the change in 
 $b_4$ splitting coefficient is more prominent than the change in $b_2$. 
 We have  estimated the solar rotation rate with varying depth and latitude. 
 In the equatorial region, the rotation first increases with depth and then 
 decreases, while an opposite behaviour is seen in the polar region. %Further 
 We also find a small but significant
  temporal variation in the rotation rate at high latitudes.   
  
\end{abstract}

\keywords{Sun:  activity -- Sun:  oscillations -- Sun:  rotation} 
\newpage
\section{INTRODUCTION} It is now well established that the frequency and splitting
coefficients of the solar {\it p}-mode oscillations vary during the sun's magnetic cycle.
The initial results were obtained from the study of the low degree modes arising
from irradiance measurements \citep{wn85} and observations from the Sun as a star
\citep{els90}.  \citet{lw90} extended this study for higher $\ell$ modes ( 5 $< \ell
<$ 140) and showed that the frequency shift increases with $\ell$ values.  On the
basis of observations made during the rising phase of cycle 22, \citet{wod91} found
a strong correlation between the frequency shift and the total surface magnetic field.
The study of solar cycle 22 was extended to the initial phase of cycle 23 by several
groups using the Global Oscillation Network Group (GONG) data and MDI data from the Solar and Heliospheric
Observatory (SOHO).  \citet{and98} analysed two 3-month equivalent data sets from
GONG and MDI during the early phase of the solar cycle 23 and found evidence of
small shifts in frequency and $a_4$ splitting coefficients.  Using the GONG data,
\citet{ab99} studied the variation in frequencies for a period of two years starting
from 1995 August.  They also presented evidence that magnetic and radiative activity indices
correlate differently with frequency shifts.  A similar result was earlier obtained
by \citet{bb93} for the declining phase of
cycle 21 and ascending phase of cycle 22.  \citet{hkh99} have also analysed GONG
frequencies over modes near 3 mHz and confirm the solar cycle changes in mean
frequencies. In addition to frequency canges, they analysed the splitting coefficients up to $a_6$ and 
reported that the even $a$
coefficients show a larger correlation with the corresponding even Legendre
component of the magnetic activity than with the average magnetic flux.  
Centroid frequencies and multiplet frequency splittings obtained from the MDI
 have also been studied by \citet{dzi98,dzi99} who primarily
concentrated on the solar cycle changes of the spherical structure of the Sun.

Recently, \citet{basu99} and \citet{hkh00} found a small temporal variation in GONG odd-order
splitting coefficients upto 15th order.  
Using different inversion techniques, they found
a systematic zonal flow migrating towards lower latitudes during the rising
phase of the cycle.  This flow can be associated with torsional oscillations as
seen in BBSO data from 1986--1990 \citep{wod93a}.  These zonal variations of the
Sun's differential rotation were first noticed by \citet{sasha97} using
MDI data.  They inferred subsurface flows from the rotational
splittings of {\it f}-mode frequencies in the range of $\ell$ from 120 to 250 using
144 days time series.  \citet{toomre} have confirmed these zonal flows using both {\it
p}- and {\it f}-mode splittings derived from MDI data.

The changes in frequencies and splittings are believed to be caused by a combination
of variations in temperature, magnetic and velocity fields.  \citet{gol91} proposed
that the variations in solar eigen-frequencies are related to the perturbations in
the surface magnetic fields.  However, \citet{kun98} argues that the frequency
changes can not be due to the variation in surface magnetic field alone, because a
reasonable explanation of the solar cycle acoustic changes must account for the
apparent photometric cycle.  

In this paper, we  analyse the frequencies and splitting coefficients obtained from the GONG project 
to extend  previous works to  all possible frequency and $\ell$ ranges.  We  also infer the solar
rotation rate as a function of depth and latitude using the  analytical
approach of \citet{mor88} involving the odd order splitting coefficients. Our results  
 reveal a significant temporal variation in the rotation rate at high latitudes over the  period  of study.

\section{FREQUENCY SHIFT AND ACTIVITY INDICES} 

	The analysis presented here uses thirty three, 108-days 
frequency 
data sets (maximum available at the time of writing) from the GONG project, covering a period of more than 3 years from 
1995 May 7 to 1998 October
17. Each data set consists of 3 GONG months (GMs), where each GM extends over 
36 days and contains $m$--averaged {\it p}-mode multiplets. The mode 
frequencies were estimated using the standard GONG analysis \citep{hil96} 
and obtained by fitting orthogonal polynomials as defined 
in  \citet{rl91}:
\begin{equation}
  \nu_{n,l,m}=\nu_{n,l}+\sum_s c_{s,n,l}\gamma_{s,l}(m) ,
\end{equation}
  where $\gamma_{s,l}$($m$) are the orthogonal polynomials for given
  value of  degree $\ell$ and
  $c_{s,n,l}$ are the splitting functions. The remaining symbols in 
  Equation~(1) have their usual meanings. Thus, this study differs from the earlier analysis of 
   GONG data by \citet{ab99} and \citet{hkh99}, wherein the frequencies used 
  were derived from    Legendre polynomial expansion.

\placetable{tb1}

The mean shift $\delta \nu$ for a given $\ell$ and $n$
is calculated from the relation 
\begin{equation}
\delta\nu(t)  =  {\sum_{n,\ell}\frac{\delta\nu_{n,\ell}(t)}{\sigma_{n,\ell}^2}}/{\sum_{n,\ell}
\frac{1}{\sigma_{n,\ell}^2}} ,
\end{equation}    
where $\delta\nu_{n,\ell}(t)$ is the change in the measured frequency for a given $\ell$ and 
radial order $n$ and 
$\sigma_{n,\ell}$ is the error in the observed frequency. The mean frequency 
shifts for GM 5--7 (1995 Sep. 23 -- 1996 Jan. 13) and GM 33--35 (1998 Jul. 2 -- 1998  Oct. 17), 
which represents low and high activity periods
 respectively, are given in Table~1 where GM 1--3 is taken as 
 the reference. It is noted that there are very few modes in the frequency range of 1500--1900 $\mu$Hz and 
3900--4300 $\mu$Hz. It is further observed that   the errors in $\delta\nu$ for these frequency ranges
are larger as compared to other ranges. 

\placefigure{fig1}
\placefigure{fig2}

The variation of $\nu$--averaged frequency shift for three different periods scaled with 
the inverse of the modal mass
as a function of $\ell$ is shown in Figure~1, 
where the frequency shift is calculated from the mean over all relevant data points.
 It is clear from the panels in Figure~1 that the shifts are constant for $\ell$ between 40 to 80 in the 
 frequency range of 1900--4300 $\mu$Hz whereas there 
 is large scatter at both low and high degrees. The temporal evolution of the mean frequency shift for 
different $\ell$ ranges  scaled by the inverse of the modal mass 
is shown in Figure~2.  Both these figures confirm
 that the frequency shift is a weak function of degree as was pointed out by  \citet{lw90}.  
 From GONG data, \citet{hkh99}   had also  found 
 no degree dependence for $\ell$ approximately between 50 to 90 around 3mHz. 

\placefigure{fig3}
\placefigure{fig4}

In Figure~3, we show the variation of $\ell$--averaged frequency shift 
as a function of frequency 
for three independent periods with reference to mean over all these periods. It is evident that
the frequency shift  has a strong 
dependence on the frequency which increases with increasing frequencies.
This is also consistent with the result of 
\citet{lw90} and \citet{ang92}. The temporal evolution of 
frequency shifts over the entire period of three and half years for different frequency 
ranges is shown in Figure~4. 
It is noted that the  frequency shifts for all frequency ranges are small upto 
GM 24--26 and then increases rapidly.  
This behaviour of the shifts 
can be understood in terms of the rapid increase of solar activity
 represented in the same figure by means of the 10.7 cm radio flux.
  It is  further noticed that the frequency shifts for all frequency ranges seem to converge 
 at GM 23--25. This particular aspect was investigated in detail using different
 combination of GMs as reference months. We conclude that this converging feature depends on reference 
frequencies and has no physical significance. 
   
The mean shift is  correlated with different solar activity indices
by computing a mean value for each index over the same 108 days interval over
which the frequencies were determined. 
The activity indices considered are: $R_I,$ the International 
sunspot number obtained from the Solar Geophysical Data (SGD); KPMI, Kitt Peak 
Magnetic Index from Kitt peak full disk magnetograms; SMMF, Stanford Mean 
Magnetic Field from SGD;  MPSI, Magnetic Plage Strength Index from Mount Wilson 
magnetograms;  FI, total flare index from SGD and T. Ata\c{c} (1999, private communication); 
He~I, equivalent width of He~I 
10830 \AA ~line, averaged over the whole disk from Kitt peak;  Mg~II, 
core-to-wing ratio of 
Mg~II line at 2800 \AA  ~from L. Floyd (1999, private communication); 
$F_{10}$, integrated radio flux at 
10.7 cm from 
SGD. A linear least square fitting is carried out between mean activity 
indices and frequency shifts.
The fitting parameters, $\chi^2$ values and both Pearson's ($r_p$) and 
Spearman's ($r_s$) correlation coefficients  
are summarised in Table~2. 
It is observed that both the magnetic and 
radiative indices are equally correlated with the frequency shifts in contrast 
to the earlier results of 
\citet{bb93} and \citet{ab99}. This aspect needs to be 
investigated in detail with the availability of more high precision data sets in future.   
 The correlation coefficients for different frequency ranges are given in Table~3. 
It is observed that the correlation has an increasing trend with frequency ranges and 
are marginally different at higher frequencies.

\placetable{tb2}
\placetable{tb3}
 
\section{SPLITTING COEFFICIENTS}

It is known that the solar differential rotation and other symmetry 
breaking factors like magnetic field can lift the degeneracy of the solar 
acoustic modes and split the eigen frequencies. Individual mode splittings
can  be represented by  Legendre polynomial expansion in ($m/L$) 
\begin{equation}
\nu_{n,\ell,m} = \nu_{n,\ell} + L \sum_{s=1}^{m} a_{s,n,\ell} P_s(m/L), 
\end{equation}
where the expansion coefficients $a_{s,n,\ell}$ are known as  the splitting coefficients
and $L^2$~=~$\ell(\ell + 1)$. 
 Alternatively, the splitting data  can be expressed in terms of 
 the splitting functions, $c_{s,n,\ell}$, as given in Equation~(1).  
  These functions 
 are generally converted  to a new set of splitting coefficients $b$'s  by using the relation \citep{rl91}
 \begin{equation}
 c_s = {\frac{(4\pi)^{1/2}}{(2\ell +1) L^2 H^1_s F_s}} b_s ,
 \end{equation}
 where 
  \begin{equation}
 H_s^1 = (-1)^{(\ell -1)} \pmatrix{ s & \ell & \ell \cr 0 & 1 & -1}, 
 \end{equation}
 and 
 \begin{equation}
  F_s = \left[ {\frac{(2\ell -s) !}{(2\ell +s +1) !}}\right]^{1/2} .
 \end{equation}
    The odd-order coefficients  measure the solar 
 rotation while the even-order coefficients  probe the  symmetry about the
 equator. The nonzero values of these even coefficients reflect the pole--equator asymmetries
 in the solar structure. It may be noted that we have retained terms up to  $s$ = 5 
  because the coefficients and their  errors are of similar magnitude for $s$~$>$~5 in the data sets 
  used in this study. 
  
\placefigure{fig5}
 
\placetable{tb4}

 The temporal variation of $b_2$ and $b_4$  coefficients is shown in Figure~5.   
 We find that $b_2$ has 
  strong correlation with activity, whereas  $b_4$ is anticorrelated. 
 The results of linear regression analysis between these coefficients and activity
 indices are given in Table~\ref{tb4}. It may be noted that $b_2$ has a better 
 quantitative correlation than $b_4$. We observe that 
 during the onset of cycle 23, the change in $b_4$ is more prominent than the
change in $b_2$.  
\citet{hkh99} have also studied the temporal variation of the $b_2$ and $b_4$ 
 coefficients at 3 mHz and reported a phase shift of one year between them. 
 The  odd order coefficients, $b_1$, $b_3$ and $b_5$, calculated 
using the expression (4) 
are used to obtain information on the variation of
the Sun's rotation rate with depth and latitude.

\subsection{Solar Rotation Rate}

The solar rotation rate using the helioseismic data are generally obtained through two different 
methods: in {\it forward approach} the frequency splittings
are computed for a chosen solar rotation model and then compared with the
observed splittings. In the {\it inverse method}  the
measured frequency splittings are used directly to produce  a single
function for the angular velocity.
In this study,  we use the analytical method of \citet{mor88} where 
the appropriate combination of odd order splitting coefficients reflects the depth variation 
of angular velocity at chosen co--latitude $\phi$, (90$^{\arcdeg}$ $-$ $\theta$,
 where $\theta$ is the latitude). 
\begin{equation}
{\Omega}^{nl}(\phi) = \sum_{i=0}^{2} d_{2i+1}(\phi)a_{2i+1}^{nl},
\end{equation}
where 
\begin{eqnarray}
d_1&=&1, \\
d_3&=&[1-5cos^2\phi], \\
d_5&=&[1-14cos^2\phi + 21 cos^4\phi] .
\end{eqnarray}
The corresponding rotation rate at equator is given by 
\begin{equation}
{\Omega}^{nl}(90^{\arcdeg}) = a_1^{nl} + a_3^{nl} + a_5^{nl} ,
\end{equation}
and at pole 
\begin{equation}
{\Omega}^{nl}(0^{\arcdeg}) = a_1^{nl} - 4a_3^{nl} + 8a_5^{nl} .
\end{equation}

\citet{wbl97} have recently modified expressions (7-12) by including
higher order splitting coefficients up to $a_7$. The addition of further terms to these 
expressions seems to be of limited utility due to the larger relative errors of higher 
order coefficients.  We compute $a$'s from $c$'s 
using Equations (61-63) of \citet{rl91}.
In Figure~6, we show the variation of $a_1$, $a_3$ and $a_5$ with degree. The non-zero values of 
these coefficients indicate that the solar rotation rate changes with depth and latitude.

\placefigure{fig6} 
\placefigure{fig7}

In Figure~7, we show the sun's rotation rate at different latitudes as
a function of depth approximated by $\nu /L$ which corresponds to the
radius of lower turning point $r_t$ as defined by the relation \citep{jcd91}
\begin{equation}
r_t = {\frac{c(r_t)}{2\pi}} {\frac{L}{\nu}}
\end{equation}
where $c$ is the sound speed. A higher value of $\nu/L$
denotes a smaller value of $r_t$ and hence a greater depth.  As a result
the low degree modes are sensitive to the rotation from surface to the
core, while the high degree modes probe solar rotation near the
surface.  From Figure~7, we find that for $\theta$ = 0$^{\arcdeg}$
and 30$^{\arcdeg}$, the rotation rate first increases and then
decreases as depth increases.  On the other hand, the rotation rate
for $\theta$ = 45$^{\arcdeg}$ and 60$^{\arcdeg}$ remains constant
below the base of the convection zone ( $\nu/L$ $\approx$ 70), beyond
which it starts increasing.  We further notice that for
$\theta$~=~$60^{\arcdeg}$, the increase in internal rotation rate
above $\nu /L$ = 70, is steeper than at $\theta$~=~$45^{\arcdeg}$.
This suggests that, as the depth increases, the equatorial rotation
rate decreases while the polar rotation rate increases.  \citet{wbl97}
have analysed frequency splittings derived from first four months of
GONG data and showed that the greatest changes in rotation occurs in
the region below the base of the convection zone.  They also found
that the data does not support models exhibiting a discontinuous shear
between the convection zone and a uniformly rotating radiative region.
\citet{thomp96} have also estimated the solar rotation rate using the
first few month's of GONG data.  Our results derived from 
analytical expressions are in gross agreement with earlier studies.

The observed rotation rate is conventionally expressed in terms of even powers of cos$\phi$:
\begin{equation} \Omega(r,\phi) = A + B~cos^2\phi + C~cos^4\phi ,
\end{equation} where $\phi$ as defined earlier is the solar co--latitude.
The surface constants $A$, $B$, and $C$ are related to splitting
coefficients by the relations %following analytical expressions
 \begin{eqnarray} 
 a_1&=& A + {\frac {1}{5}} B + {\frac {3}{35}} C ,\\ 
 a_3 &= & - {\frac{1}{5}} B - {\frac {2}{15}} C ,\\ 
 a_5 &= &{\frac {1}{21}} C .
\end{eqnarray}

Various workers have calculated these surface coefficients using
different data sets.  \citet{brw89} found $A$ = 462.8 nHz, $B$ =
$-$~56.7 nHz and $C$ =~$-$ 75.9 nHz for $r~\ge~$0.723~$R_\odot$ using
CaII K intensity data taken from South pole.  Based on 100-day
observations made at BBSO for $\ell$ between 10 to 60, \citet{lib89}
found the best fit with $A$ = 461~nHz, $B$ = $-$ 60.5 nHz and $C$ =
$-$ 75.4 nHz.  We have derived these constants from the GONG data for the 
20 $\le$ $\ell$  $\le$ 100 and
obtained the best fit with $A$= 459.54 nHz, $B$ = $-$~61.90 nHz and
$C$ = $-$~70.20 nHz. Using the odd-order splitting coefficients of GONG data 
derived from Legendre polynomial expansion, 
we obtain, $A$ = 459.97 nHz, $B$~=~$-$~61.33 nHz, and 
$C$ = $-$~70.93 nHz.  The values obtained for $A$, $B$ and $C$ using GONG data 
are in close agreement with the earlier values.

\placefigure{fig8}

In Figure~8, we plot the average surface rotation rate as a function
 of latitude using the derived coefficients. In the same plot, we have also shown the results from
 inversion techniques (Antia, private communication) and Doppler surface 
 measurements \citep{snod84}.
 It  is clear that the rotation rate changes significantly in mid latitude
 while the change in rotation rate near pole and  equator is
 small. We find that the rotation rate derived from the GONG data agrees 
 well with other results. However, the inverted rotation profile 
 departs from other rotation rates beyond the latitude of 70$^{\arcdeg}$, probably due to the 
 resolution limitation in inversion techniques. 
 
 \placefigure{fig9}
 
 \citet{wod93} studied the time variation of equatorial rotation rate for
different $\ell$ ranges and suggested a small variation from year to year.
To investigate temporal variations, Figure~9 shows the residual surface rate at four different latitudes, 
obtained after subtracting the average angular velocity shown in Figure~8 by long dash line. It is evident that at higher 
latitudes, the residual rotation rate, commonly known as zonal flows, is time dependent. This has a magnitude of 
approximately 2 nHz at a latitude of 60$^{\arcdeg}$. Recently \citet{basu99}, \citet{hkh00} using GONG data, and 
  \citet{toomre} using  MDI data, have also reported a small but significant time variation in the rotation rate. 
 Our results derived from the analytical method are  consistent with these inversion studies. 
\section{CONCLUSION}

The mean frequency shift varies over the solar cycle and is correlated
similarly with both the magnetic and radiative indices.  

The $b_2$ splitting coefficient is linearly correlated with the
activity indices, while $b_4$ is anti-correlated for all degree and 
frequency ranges. 

We  detect a small but significant variation in the
rotation rate derived from the linear combination of odd order
coefficients over a period of three and half years.

\begin{acknowledgements} 
We thank the anonymous referee for his critical remarks which impreved the manuscript.
 We also thank T. Ata\c{c}, L. 
Floyd, and R.K. Ulrich for supplying us the Flare index, Mg II, and MPSI data 
respectively and H. M. Antia for providing us the inversion results. 
This work utilises data obtained by the Global Oscillation Network Group project, managed by the National Solar 
Observatory, a Division of the National Optical Astronomy Observatories, which is operated by 
AURA, Inc. under cooperative agreement with the National Science Foundation. The data were 
acquired by instruments operated by Big Bear Solar Observatory, High Altitude Observatory, 
Learmonth Solar Obsrvatory, Udaipur Solar Observatory, Instituto de Astrophsico de Canaris, and 
Cerro Tololo Interamerican Observatory. NSO/Kitt Peak magnetic, and Helium measurements used 
here are produced cooperatively by NSF/NOAO; NASA/GSFC and NOAA/SEL.  This work is 
partially supported under the CSIR Emeritus Scientist Scheme and Indo--US collaborative 
programme--NSF Grant INT--9710279. 
\end{acknowledgements}

\clearpage

\figcaption[fig1.eps]{The binned frequency shifts 
for three different GONG months, scaled by the inverse of the modal mass, 
 as a function of degree $\ell$. The shifts have been calculated 
 with respect to the mean of all three  periods. 
 Error bars represent mean error in shift.\label{fig1}}

\figcaption[fig2.eps]{The temporal evolution of the frequency shifts scaled by the inverse of the modal mass 
 for three different $\ell$ ranges: 20--40 (traingles),
 40--80 (stars) and 80--100 (diamonds). The corresponding period 
is given in the upper x-axis. \label{fig2}}

\figcaption[fig3.eps]{The binned frequency shifts for three 
 independent periods: GONG months 8--10 (squares), 23--25 (diamonds) and 
33--35 (triangles) as a function of frequency. The shift has been calculated with 
reference to the mean of all three periods.  Error bars represent mean error in shift.\label{fig3}}

\figcaption[fig4.eps]{The temporal evolution of frequency shifts for different frequency ranges:
1900--2300 $\mu$Hz (plus), 2300--2700 $\mu$Hz (stars), 2700--3100 $\mu$Hz (squares), 3100--3500 $\mu$Hz 
(dimonds), and 3500--3900 $\mu$Hz (triangles). The  solid line without symbols represents the total 
mean shift in the range of 1900--3900 $\mu$Hz.  The shifts have been 
calculated with respect to the mean of all 33 data sets. The solid line with crosses 
represents the scaled 10.7 cm radio flux.\label{fig4}}

\figcaption[fig5.eps]{The temporal evolution of $b_2$ (solid line) and $b_4$ (dashed line)
coefficients derived from the splitting functions $c_s$.  
The solid line with diamonds represents the scaled 10.7 cm radio flux.
\label{fig5}}

\figcaption[fig6.eps]{ The variation of splitting coefficients with degree $\ell$ for GM 8--10. The solid line
represents the average values of the coefficients for $\ell$ = 20--100. The 1$\sigma$ errors 
are shown by dotted line.\label{fig6}}

\figcaption[fig7.eps]{ The solar rotation rate 
as a function of $\nu/L$, a proxy for the turning point radius $r_t$ (Equation~13), 
for five different latitudes: $\theta$=0$^{\arcdeg}$ (plus), $\theta$=30$^{\arcdeg}$ (stars),  
$\theta$=45$^{\arcdeg}$ (diamonds), 
$\theta$=60$^{\arcdeg}$ (triangles), and $\theta$=90$^{\arcdeg}$ (squares) are shown. 
%The errors are of similar magnitude as shown in Figure~7. 
\label{fig7}}

\figcaption[fig8.eps]{The variation of average solar rotation rate calculated from three term fits 
in even power of cos$\phi$ are shown as a function of latitude. The solid line represents
 the rotation rate from \citet{snod84}, dotted and dash-dot-dot-dot lines are for BBSO data sets, 
 as derived by \citet{brw89} and \citet{lib89} respectively.  
 Other lines represent rotation rates calculated from GONG data;  
 long dash  line for analytical results  using orthogonal polynomials,  
and short dash for inversion results (Antia,  private communication). \label{fig8}}

\figcaption[fig9.eps]{The temporal evolution of residual rotation rate  near 
the surface at four different latitudes.  
\label{fig9}}

\clearpage
\begin{deluxetable}{crrr} 
\tablecolumns{4} 
\tablecaption{NUMBER OF COMMON MODES AND FREQUENCY SHIFTS FOR  $\ell$ BETWEEN 20 TO 100 
AND DIFFERENT $\nu$ RANGES \label{tb1}}
\tablewidth{0pt}
\tablehead{
\colhead{Frequency range} & \multicolumn{1}{c}{Number} &\multicolumn{2}{c}{Frequency shifts}\\  
\colhead{($\mu$Hz)}& \colhead{of modes}&\multicolumn{2}{c}{(nHz)}\\
\cline{3-4} 
\colhead{}&\colhead{}&\colhead{GM 5--7}&\colhead{GM 33--35}} 
%\colhead{}&\colhead{}&\multicolumn{2}{c}{GM 5--7} {GM 33--35}}
\startdata
1500 $\le \nu \le$ 1900 & 6 &  11.4 $\pm$ 5.51 & 32.2 $\pm$ 6.12 \\
1900 $\le \nu \le$ 2300 & 71 & $-$04.4 $\pm$ 1.71 & 40.9 $\pm$ 1.86\\
2300 $\le \nu \le$ 2700 & 92 & $-$05.8 $\pm$ 1.64& 96.5 $\pm$ 1.82\\ 
2700 $\le \nu \le$ 3100 & 66 & $-$16.7 $\pm$ 1.71 & 173.7 $\pm$ 2.06\\ 
3100 $\le \nu \le$ 3500 & 84 & $-$13.4 $\pm$ 1.99 & 270.4 $\pm$ 2.39\\ 
3500 $\le \nu \le$ 3900 & 110 & $-$29.3 $\pm$ 3.63 & 338.3 $\pm$ 4.05 \\ 
3900 $\le \nu \le$ 4300 & 31 & $-$34.5 $\pm$ 16.4 & 377.5 $\pm$ 17.6 \\
1900 $\le \nu \le$ 3900  & 423 & $-$10.4 $\pm$ 0.08 & 139.1 $\pm$ 0.10\\
\enddata
\end{deluxetable} 
\clearpage
\begin{deluxetable}{lccccc} 
\tablecaption{FITTING AND CORRELATION STATISTICS IN THE FREQUENCY RANGE
OF 1900--3900 $\mu$Hz AND $\ell$ BETWEEN 20 TO 100\label{tb2}}
\tablewidth{0pt}
\tablehead{
\colhead{Activity index}
& \colhead{slope} & \colhead{intercept} & \colhead{$\chi^2$} & \colhead{$r_p$} & 
\colhead{$r_s$}\\
\colhead{}&\colhead{(nHz per activity unit)}&\colhead{(nHz)}&\colhead{}&\colhead{}&\colhead{}
}
\startdata

$R_I$&2.154 $\pm$ 0.07& $-$52.802 $\pm$ 0.24 & 14.0&0.98 & 0.93 \\
KPMI& 25.04 $\pm$ 0.90 & $-$205.67 $\pm$ 0.74  &26.6&  0.96 & 0.82 \\
$F_{10}$&2.758 $\pm$ 0.09& $-$230.84 $\pm$ 0.83  &10.3& 0.99 & 0.95 \\
He I & 8.416 $\pm$ 0.30 & $-$391.20 $\pm$ 1.41  & 31.5&0.95 & 0.92 \\
SMMF & 7.283 $\pm$ 0.36 & $-$85.356 $\pm$ 0.44  &397& 0.87 & 0.69 \\
MPSI&136.6 $\pm$ 4.87& $-$48.772 $\pm$ 0.23&9.20&0.99&0.92\\
FI &27.69 $\pm$ 1.58 & $-$39.642 $\pm$ 0.22 &50.4&  0.93 & 0.88 \\
Mg II& 19580 $\pm$ 697 & $-$5001.6 $\pm$ 17.8 &7.30&  0.99 & 0.93\\
\enddata

\end{deluxetable} 
\clearpage
\begin{deluxetable}{lcccccccccc} 
\tablecaption{CORRELATION STATISTICS FOR DIFFERENT FREQUENCY RANGES\tablenotemark{*}
\label{tb3}}
\tablewidth{0pt}
\tablehead{
\colhead{Activity index}
& \multicolumn{2}{c}{1900--2300} &\multicolumn{2}{c}{2300--2700 }&\multicolumn{2}{c}{2700--3100}
&\multicolumn{2}{c}{3100--3500}&\multicolumn{2}{c}{3500--3900 }\\
\colhead{}&\colhead{$r_p$} & \colhead{$r_s$}&\colhead{$r_p$} & \colhead{$r_s$}&\colhead{$r_p$} & \colhead{$r_s$}&
\colhead{$r_p$} & \colhead{$r_s$}&\colhead{$r_p$} & \colhead{$r_s$}
}
\startdata

$R_I$&0.92 & 0.86&0.97&0.88&0.98&0.92&0.99&0.95&0.98&0.93 \\
KPMI&  0.89 & 0.76&0.94&0.74&0.96&0.81&0.98&0.89&0.98&0.88 \\
$F_{10}$& 0.91 & 0.89&0.97&0.89&0.98&0.95&0.99&0.97&0.99&0.96 \\
He I & 0.84 & 0.86&0.93&0.84&0.95&0.90&0.97&0.97&0.97&0.97 \\
SMMF &  0.70 & 0.69 &0.77&0.68&0.80&0.67&0.82&0.68&0.82&0.72\\
MPSI&0.93&0.87&0.98&0.86&0.99&0.92&0.99&0.95&0.99&0.96\\
FI & 0.84 & 0.84&0.91&0.85&0.92&0.88&0.94&0.90&0.94&0.87\\
Mg II&   0.92 & 0.86&0.98&0.86&0.99&0.93&0.99&0.97&1.00&0.96\\
\enddata
\tablenotetext{*}{Frequencies are in $\mu$Hz}
\end{deluxetable} 
\clearpage
\begin{deluxetable}{lccccc} 
\tablecaption{FITTING AND CORRELATION STATISTICS FOR  $b_2$ AND $b_4$ COEFFICIENTS\label{tb4}}
\tablewidth{0pt}
\tablehead{

\colhead{Activity}&\colhead{slope} & \colhead{intercept}
&\colhead{$\chi^2$}
 &\colhead{$r_p$}&\colhead{$r_s$} \\ 
 \colhead{index}&\colhead{(nHz per activity unit)}&\colhead{(nHz)}&\colhead{}&\colhead{} \\ 
 \cline{1-6} 
 \colhead{} &\colhead{}&\colhead{$b_2$} &\colhead{}&\colhead{}&\colhead{}}
\startdata
$R_I$&10.26 $\pm$ 3.09& $-$221.48 $\pm$ 93.60 & 1.0 & 0.92 & 0.80 \\
KPMI& 122.28 $\pm$ 36.19 & $-$973.70 $\pm$ 298.10 &0.6   & 0.95 & 0.82 \\
$F_{10}$&13.22 $\pm$ 3.94 & $-$1076.61 $\pm$ 330.48 &0.8 & 0.93 & 0.81 \\
He I & 40.96 $\pm$ 12.21 & $-$1873.15 $\pm$ 565.20  &0.8 & 0.94 & 0.88 \\
SMMF & 34.80 $\pm$ 14.60 & $-$378.15 $\pm$ 174.85 &6.3 & 0.88 & 0.75 \\
MPSI&656.86 $\pm$ 195.16& $-$204.22 $\pm$ 89.24& 0.7 & 0.93&0.80\\
FI &117.50 $\pm$ 63.70 & $-$149.81$\pm$ 88.81 & 0.7 &  0.91 & 0.81 \\
Mg II& 94543.52 $\pm$ 27952.12 & $-$24119.43 $\pm$ 7134.69 & 0.6 & 0.94 & 0.85\\
\cline{1-6}
\colhead{} &\colhead{}&\colhead{$b_4$} &\colhead{}&\colhead{}&\colhead{}\\
\cline{1-6}  
$R_I$& $-$7.45 $\pm$ 2.98 & 235.56 $\pm$ 90.23 & 1.8 & $-$0.87 & $-$0.57 \\
KPMI& $-$81.62 $\pm$ 34.90 & 742.04 $\pm$ 287.49 & 2.6   & $-$0.79 & $-$0.42 \\
$F_{10}$& $-$9.51 $\pm$ 3.80 & 866.84 $\pm$ 318.50 & 1.8 & $-$0.88& $-$0.61 \\
He I &$-$28.56$\pm$ 11.78 & 1398.59 $\pm$ 545.15  &2.2 & $-$0.83 & $-$0.52 \\
%TSI & $-$589.17$\pm$ 256.16 & 804815.37 $\pm$ 3498.70 & 0.8 &$-$0.93 & $-$0.84 \\
SMMF & $-$24.96 $\pm$ 14.09 & 363.76 $\pm$ 168.69 &5.0 & $-$0.57 & $-$0.32 \\
MPSI&$-$467.32 $\pm$ 188.14 & 237.85 $\pm$ 85.97 & 1.9 & $-$0.89& $-$0.64\\
FI &$-$116.72 $\pm$ 61.52 & 220.35$\pm$ 85.60 &2.5 &  $-$0.77 & $-$0.56 \\
Mg II& $-$65912.68 $\pm$ 26949.38 & 16907.71 $\pm$ 6878.67 &2.1 & $-$0.86 & $-$0.54\\
\enddata
\end{deluxetable} 


\begin{thebibliography}{}
\bibitem[Anderson, Howe, \& Komm(1998)]{and98} Anderson, E. R., Howe, R.,~\& Komm, R.  1998, 
in Structure  and Dynamics of the Sun
and Sun-like Stars,  ed. S. G. Korzennik \& A. Wilson (ESA-SP-418, Noordwijk: ESA), 901
\bibitem[Anguera Gubau et al.(1992)]{ang92}Anguera Gubau, M., Pall\'{e}, P. L., P\'{e}rez Hern\'{a}ndez, F., 
R\'{e}gulo, C.,~\& Roca Cort\'{e}s, T.  1992, \aap,   255, 363
\bibitem[Bachmann \& Brown(1993)]{bb93}Bachmann, K. T.,~\& Brown, T. M.  1993, \apj, 411, L45
\bibitem[Basu \& Antia(1999)]{basu99} Basu, S.,~\& Antia, H. M. 1999,  IAU Colloquium 179
\bibitem[Bhatnagar, Jain, \& Tripathy(1999)]{ab99}Bhatnagar, A., Jain, K., 
\& Tripathy, S. C.  1999, \apj,    521, 885
\bibitem[Brown et al.(1989)]{brw89} Brown, T.M., Christensen-Dalsgaard, J., Dziembowski, W. A., Goode, P. R.,
Gough, D.O.,~\& Morrow, C. A. 1989, \apj, 343, 526
\bibitem[Christensen-Dalsgaard \& Berthomieu(1991)]{jcd91} Christensen-Dalsgaard, J., \& Berthomieu, G., 1991, 
in Solar Interior and Atmosphere, ed. A. N. Cox, W. C. Livingston, \& M. S. Matthews (Tucson: Univ. Arizona Press), 401
\bibitem[Dziembowski et al.(1998)]{dzi98}Dziembowski, W. A., Goode, P. R., DiMauro, M. P.,
Kosovichev, A. G.,~\& Schou, J.  1998, \apj,    509, 456
\bibitem[Dziembowski et al.(1999)]{dzi99}Dziembowski, W. A., Goode, P. R.,
Kosovichev, A. G.,~\& Schou, J.  1999, \apj, submitted
\bibitem[Elsworth et al.(1990)]{els90}Elsworth, Y., Howe, R., Isaak, G. R., McLeod, C. P.,~\& 
New, R. \nat, 345, 322    
\bibitem[Goldreich et al.(1991)]{gol91}Goldreich, P., Murray, N., Willette, G.,~\& 
Kumar, P.  1991, \apj,    370, 752
\bibitem[Hill et al.(1996)]{hil96} Hill, F., et al.  1996, Science    272, 1292
\bibitem[Howe, Komm, \& Hill(1999)]{hkh99}Howe, R., Komm, R. \& Hill, F.  1999, 
\apj, 524, 1084
\bibitem[Howe, Komm, \& Hill(2000)]{hkh00}Howe, R., Komm, R. \& Hill, F.  2000, 
\solphys, in press 
\bibitem[Kosovichev \& Schou(1997)]{sasha97} Kosovichev, A. G., \& Schou, J., 1997, \apjl, 482, 207
\bibitem[Kuhn(1998)]{kun98} Kuhn, J. R. 1998, in Structure  and Dynamics of the Sun
and Sun-like Stars, ed. S. G. Korzennik \& A. Wilson ( ESA-SP-418, Noordwijk: ESA), 871
\bibitem[Libbrecht(1989)]{lib89} Libbrecht, K. G. 1989, \apj, 336, 1092
\bibitem[Libbrecht \& Woodard(1990)]{lw90} Libbrecht, K. G.,~\& Woodard, M. F.  1990, \nat, 345, 779
\bibitem[Morrow(1988)]{mor88} Morrow C. A. 1988, in Seismology of the Sun and 
Sun like Stars,  ed. E. J. Rolfe (ESA-SP-286, Noordwijk: ESA), 137
\bibitem[Ritzwoller \& Lavely(1991)]{rl91} Ritzwoller, M. H.,~\&  Lavely, E. M.  1991, \apj,    369, 557
\bibitem[Snodgrass(1984)]{snod84} Snodgrass, H. B.  1984, \solphys, 94, 13    
\bibitem[Thompson et al.(1996)]{thomp96} Thompson, M. J., et al.  1996, Science,   272, 1300
\bibitem[Toomre et al.(2000)]{toomre}Toomre, J., Christensen-Dalsgaard, J., Howe, R., 
Larsen, R. M., Schou, J., \& Thompson, M. J., \solphys, 2000, in press
\bibitem[Wilson, Burtonclay \& Li(1997)]{wbl97} Wilson, P. R., Burtonclay, D.,~\& Li, Y.  1997, \apj,    489, 395
\bibitem[Woodard et al.(1991)]{wod91} Woodard, M. F., Kuhn, J. R., Murray, N.,~\& Libbrecht, K. G.  1991, \apj,    373, L81
\bibitem[Woodard and Libbrecht(1993a)]{wod93a} Woodard, M. F., ~\& Libbrecht, K. G.  1993a, Science, 260, 1778
\bibitem[Woodard and Libbrecht(1993b)]{wod93} Woodard, M. F., ~\& Libbrecht, K. G.  1993b, \apj,    402, L77
\bibitem[Woodard and Noyes(1985)]{wn85}Woodard, M. F.,~\& Noyes, R. W.  1985, \nat,     318, 449

\end{thebibliography}
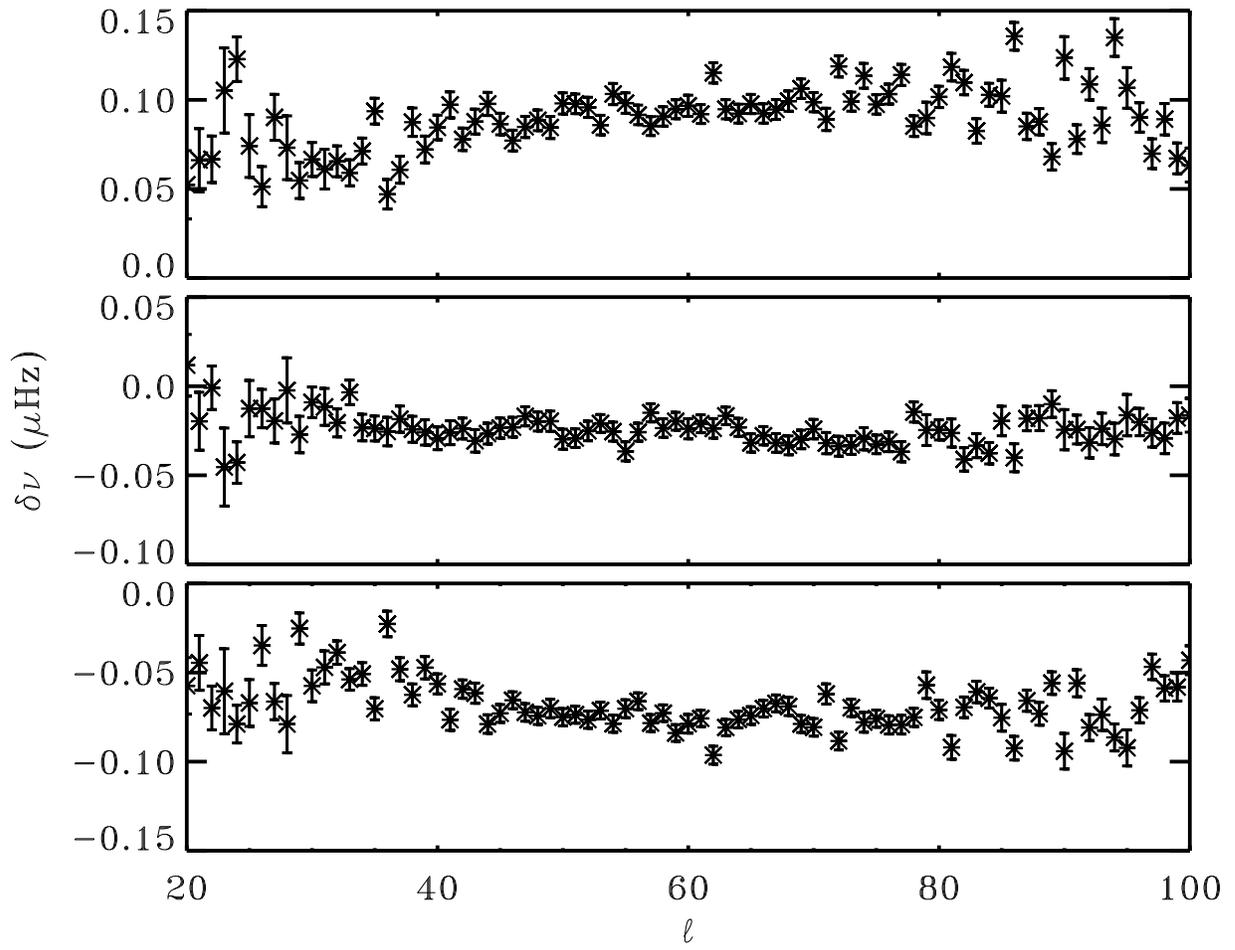
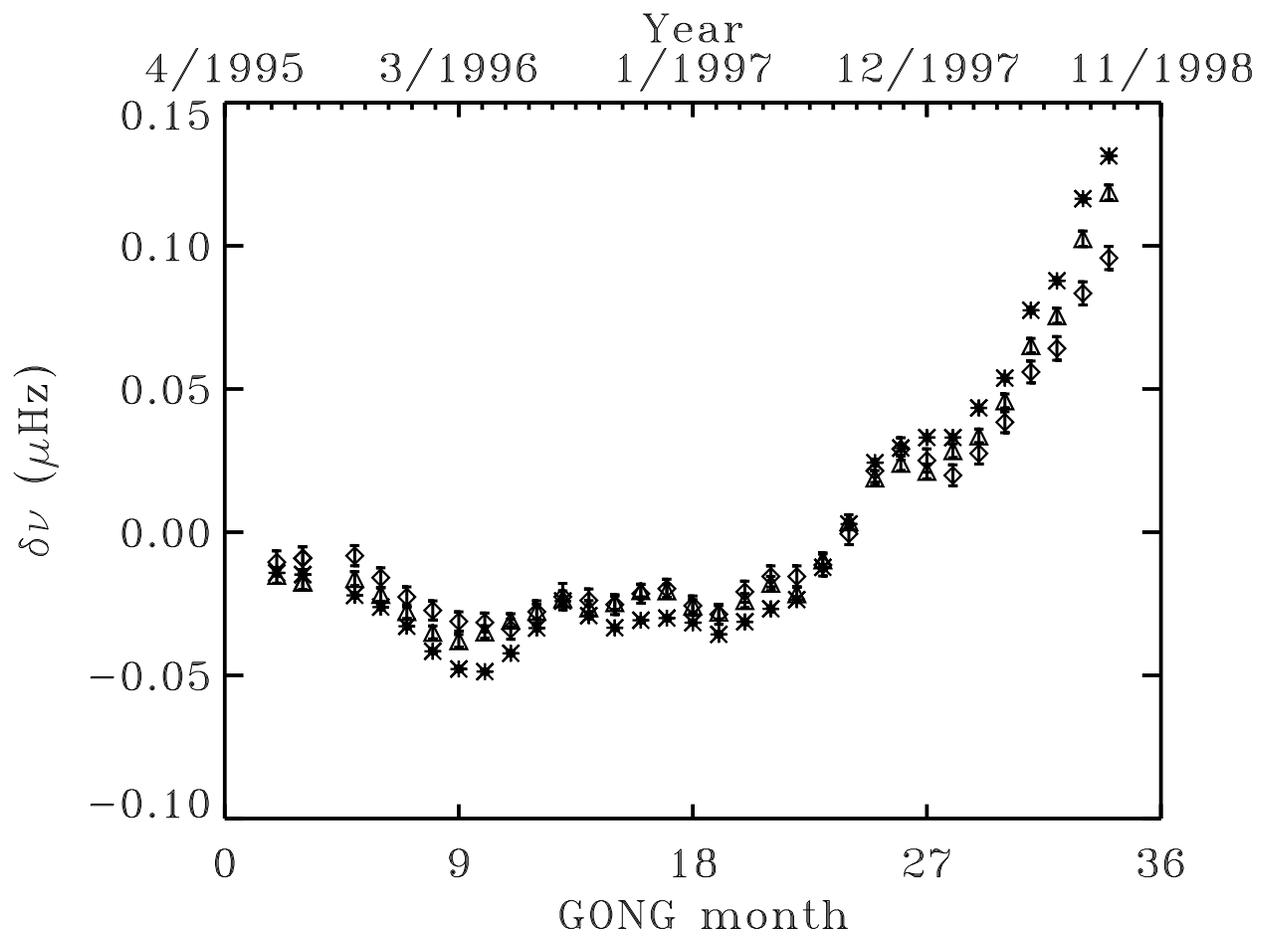
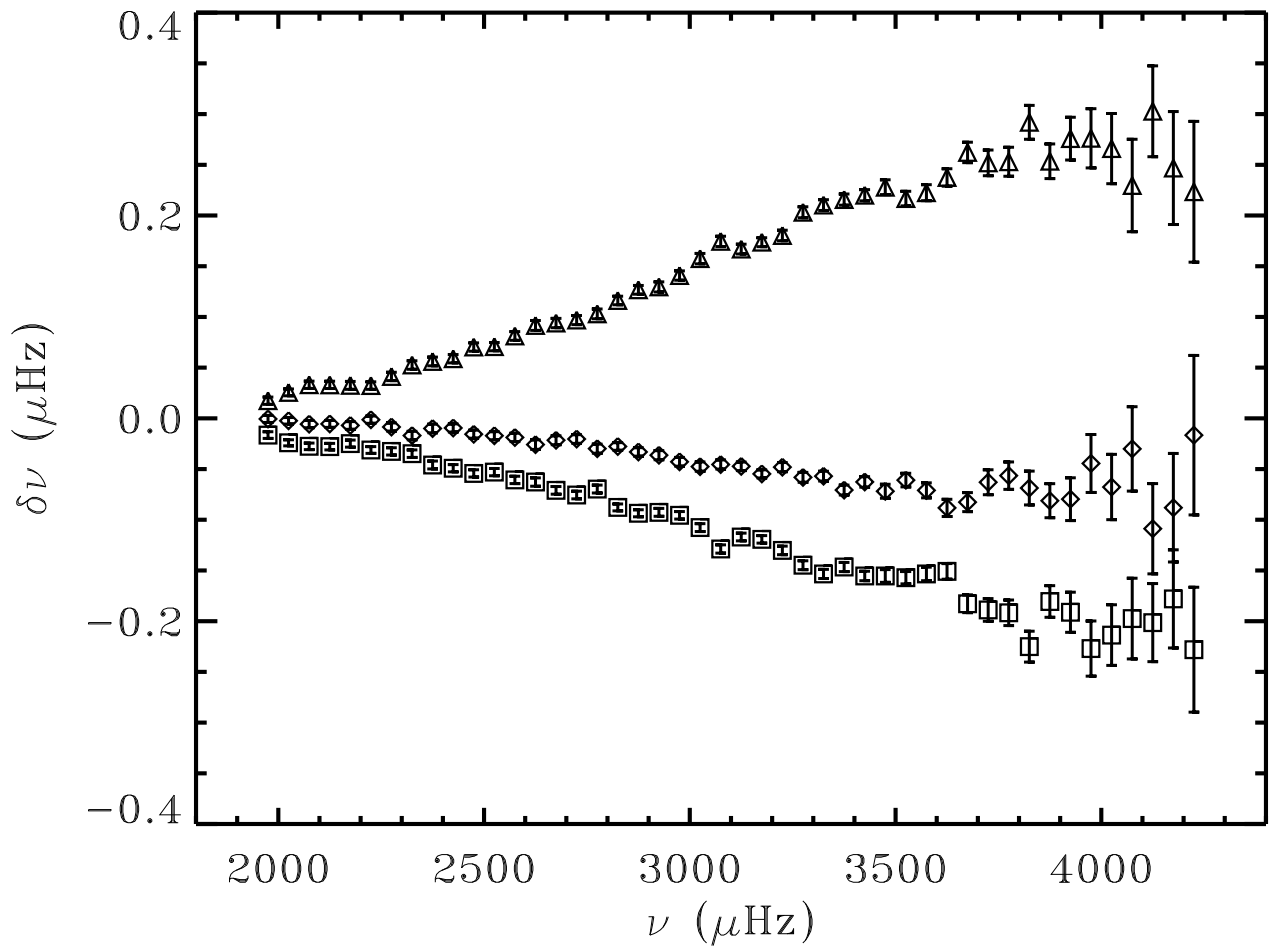
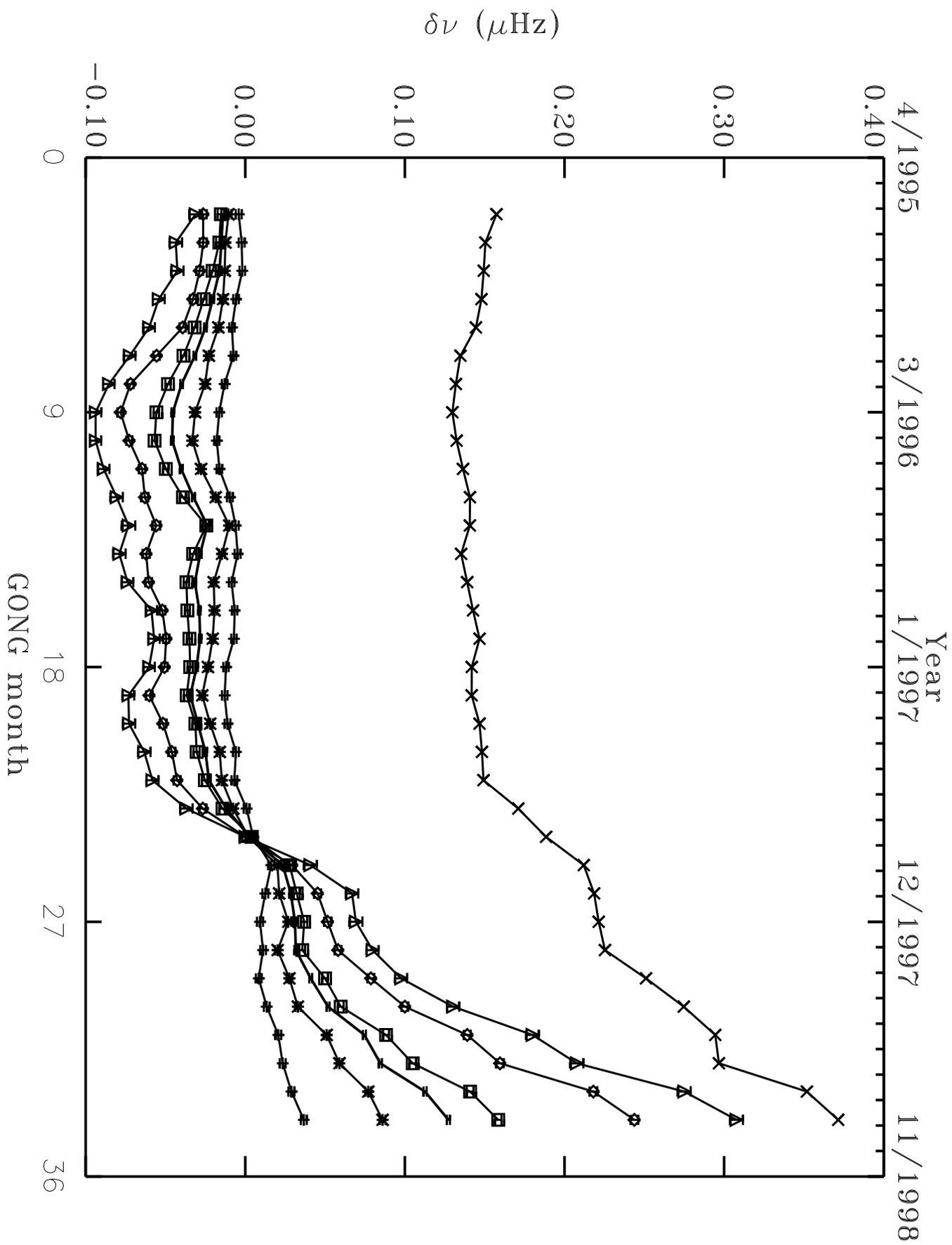
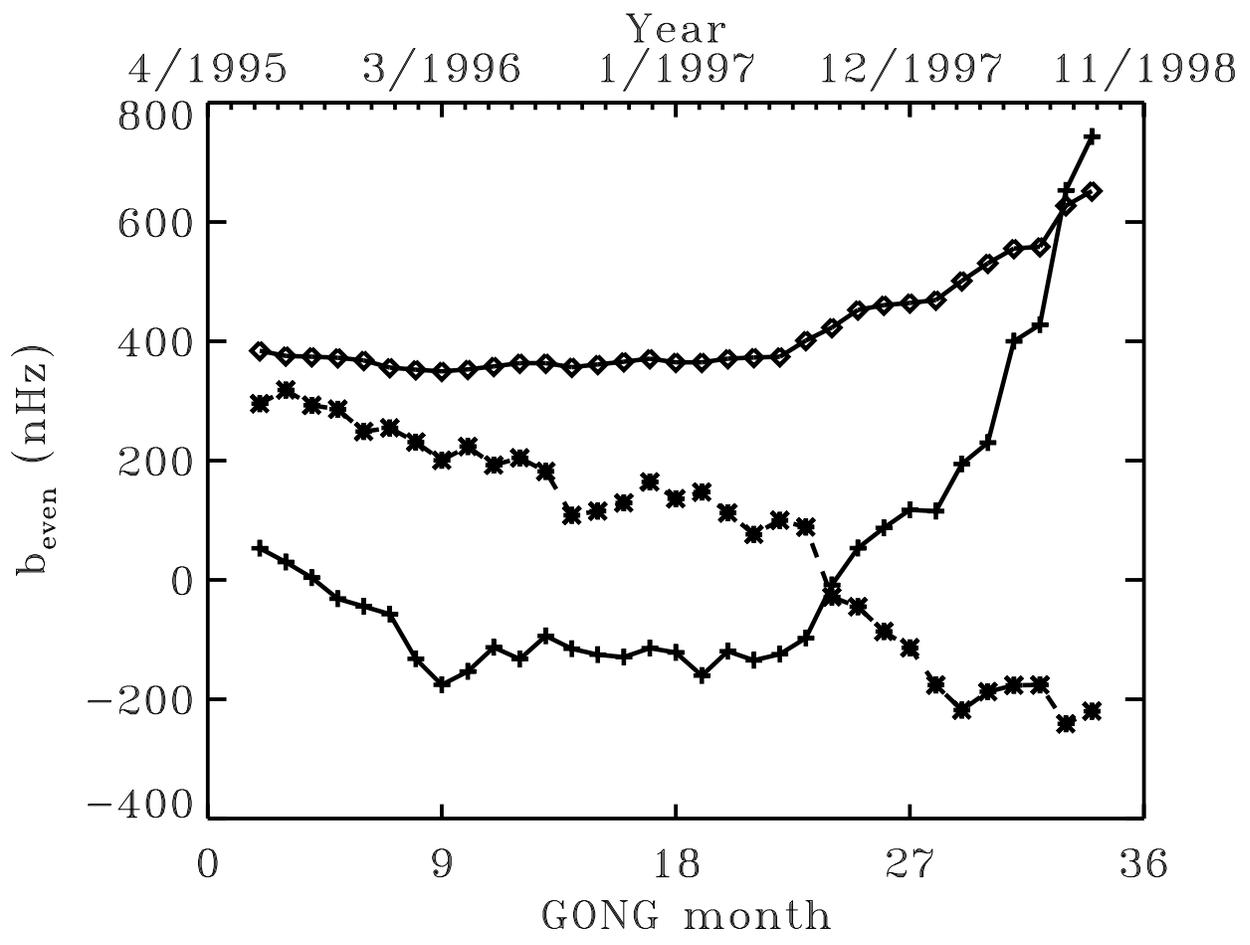
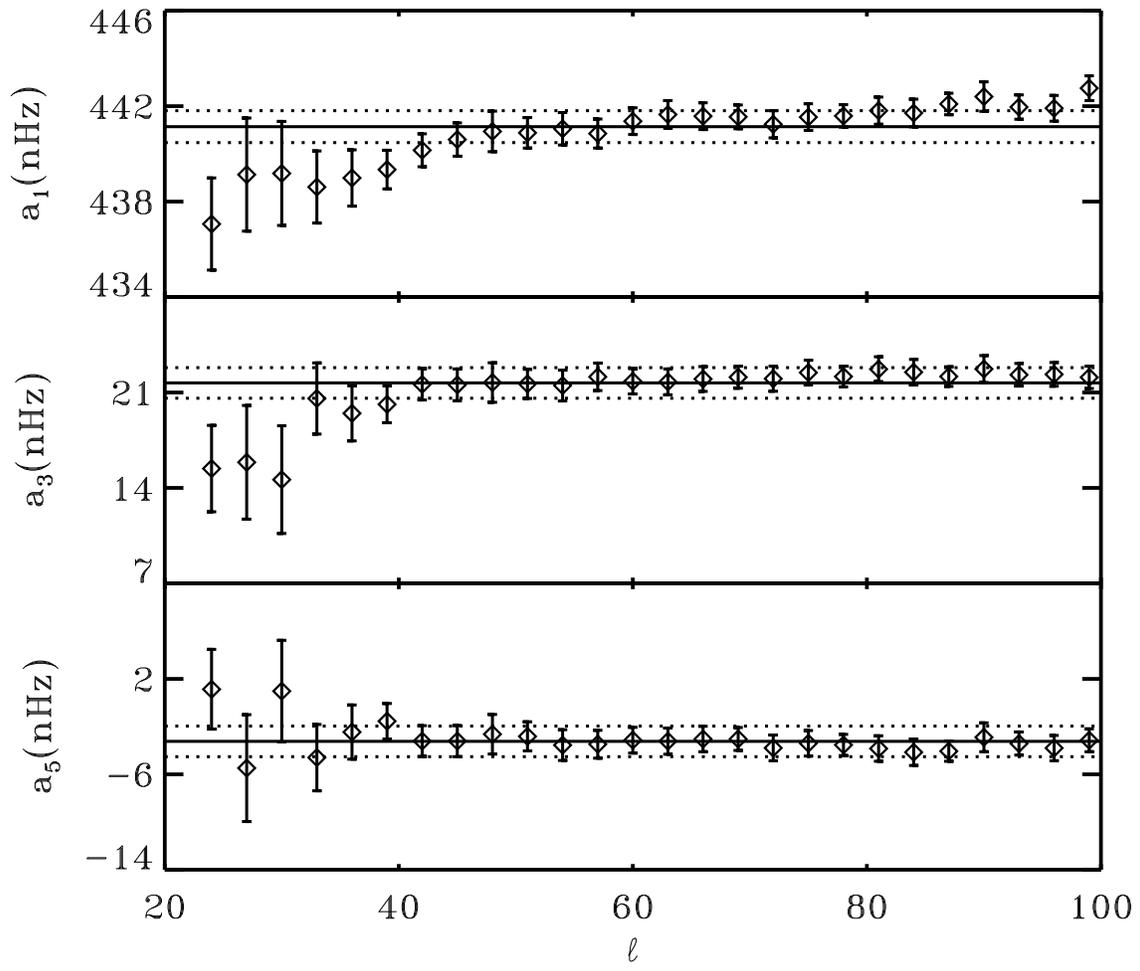
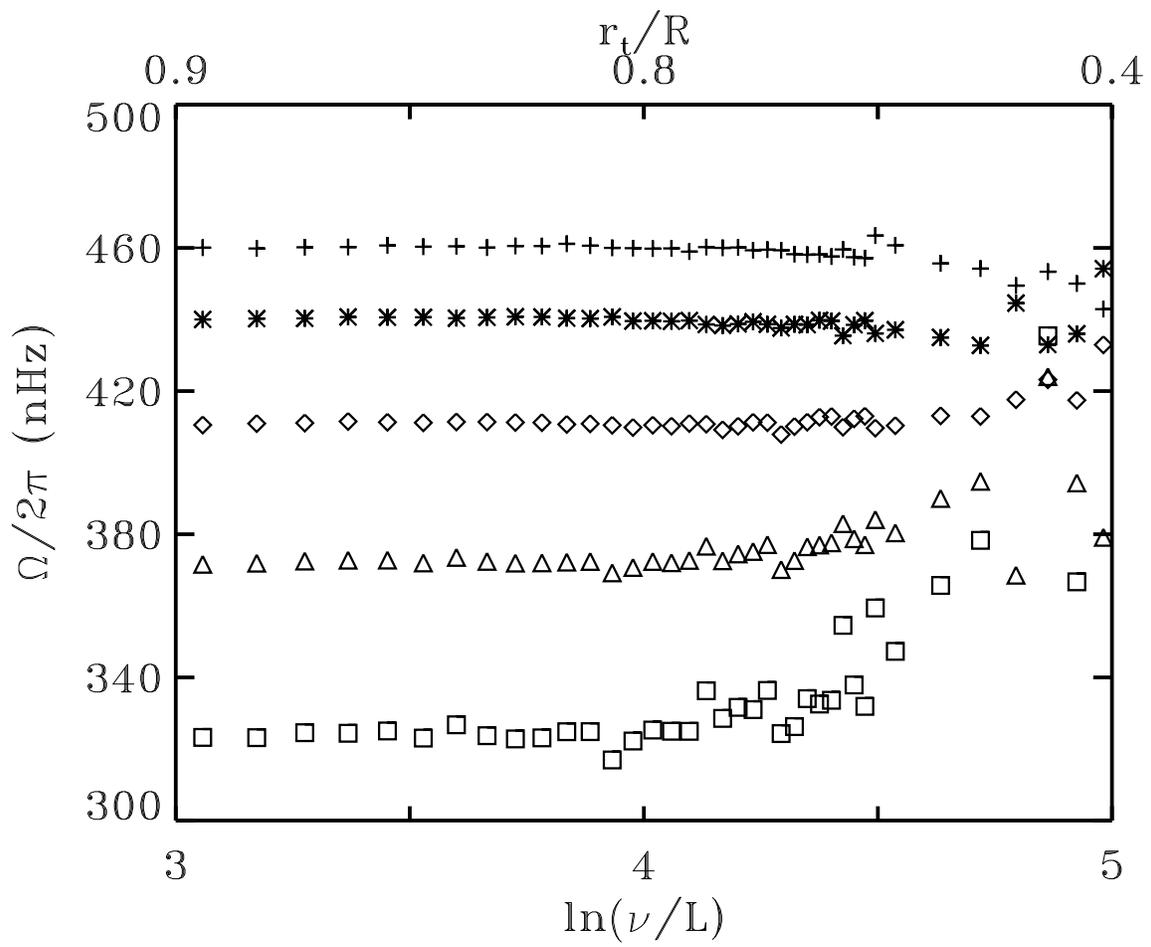
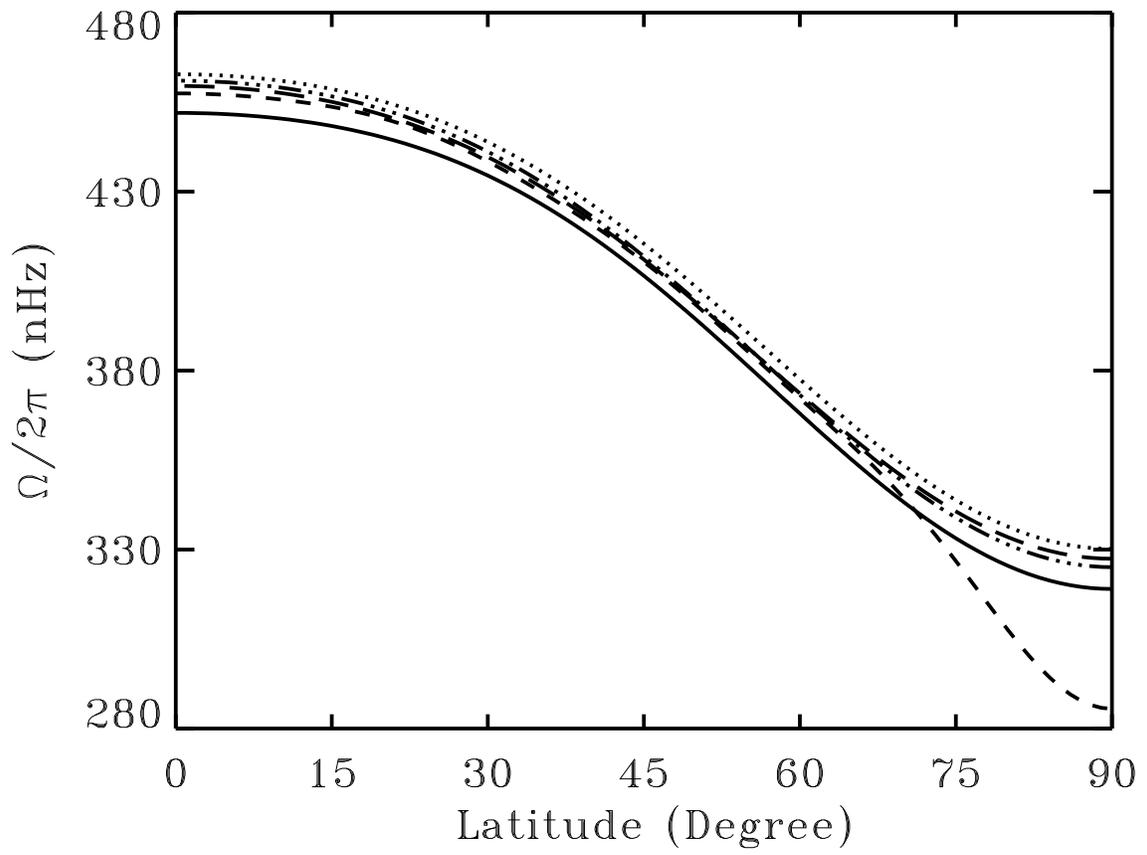
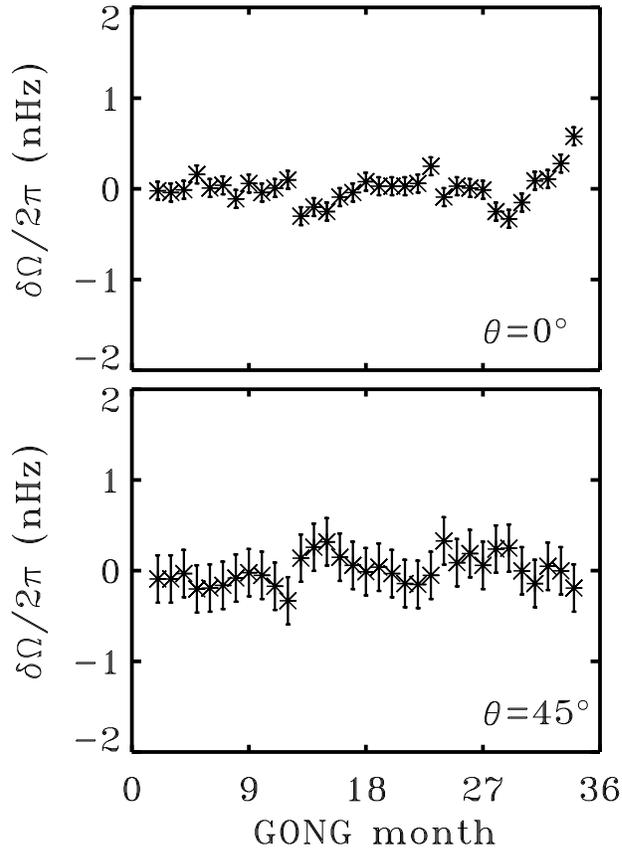
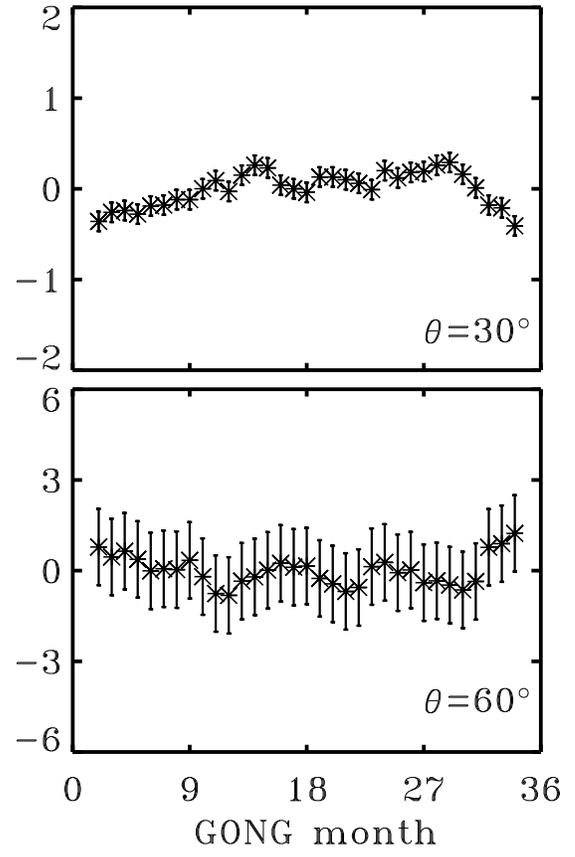
\end{document}